# A Complete Set of Grad's Thirteen Regularized Moment Equations

L. Pekker[*] and O. Pekker

ERC Inc. Huntsville, AL 35805, USA

V. Timchenko

University of New South Wales, Sydney, Australia

## ABSTRACT

This paper derives transport equations for medium rarefied gases from the Bhatnagar-Gross-Krook (BGK) model kinetic equation using a Hermite polynomial approximation for the monoatomic gas distribution function. We apply the Chapman-Enskog regularization method to Grad's velocity distribution function that corresponds to his thirteen moment equation, extending these equations to the third order of the Knudsen number. We show that the obtained set of Grad's regularized thirteen moment equations is complete, and that all previously obtained closures for Grad's thirteen moment equations are the sets of truncated equations obtained in this paper. Using the truncated transport equations may lead to losing accuracy in computing gas flows and heat transfer in rarefied gases. The velocity distribution function for the resulting 13 regularized moment equations is presented.

## I. INTRODUCTION

One of the hardest problems in computational fluid dynamics is the modeling of medium rarefied gases with the Knudsen number in the range of 0.01 – 1. In this case, the gas is rarefied to such a degree that using the Navier-Stokes-Fourier equations is questionable, and it is not sufficiently rarefied that using Direct Simulation Monte Carlo (DSMC) methods is effective. It should be stressed that in recent years modeling of flows for Knudsen numbers in this range have become more and more important for many practical applications, ranging from the modeling of re-entry of space vehicles into the atmosphere to modeling of microscale flows and heat transfer in microchannels. One of the ways to cover this range of Knudsen numbers is to use thirteen (or more) moment equations instead of the five Navier-Stokes-Fourier equations, which are the first order of the Knudsen number.

[*] Corresponding author is Leonid Pekker, Tel. +1 (661) 822-7064
 E-mail address: Leonid.Pekker@gmail.com

In 1949, Grad [1, 2] derived the 13 moment equations corresponding to the second order of the Knudsen number. Unfortunately, Grad's moment equations sometimes produce unphysical solutions; for example, they fail to describe smooth shock structures for Mach numbers above a critical value [3]. In 2003, Struchtrup and Horrihon [4] regularized Grad's 13 moment equations, extending them to the third order of the Knudsen number. The authors have developed a new closure method which principally differs from the well-known Chapman-Enskog method [5, 6] (which was used to derive a closure of Euler's gas dynamics equations), in which they have used 26 non-Hermite polynomials representation of the velocity distribution function and, corresponding, 26 moments. It should be noted that the Struchtrup-Torrihon closure method is very complicated; this is probably one of the reasons that this method and equations are difficult to comprehend. In a recent paper [7], the author suggested another closure for Grad's 13 moment equations by using 29-term Hermite polynomial approximation for monatomic gas distribution function, and applying the Chapman-Enskog regularization method to Grad's velocity distribution function that corresponds to his 13 moment equations. In their paper, the collision term in the Boltzmann equation is assumed to be in BGK form. The sets of tranport equations obtained in works [4] and [7] are different because the authors use different bases for approximation of the velocity distribution function which are non-congruent to each other.

In the present work, we sugest a new closure for Grad's thirteen moment equations in which we use the regularization method [7], applying it to the 35-term Hermite representation of the velocity distribution function, and the collision term in the BGK form. We explicitly show why this representation of the velocity distribution function is complete in terms of expansion of Grad's thirteen moment equations to third order of the Knudsen number, and therefore, has good physical sense. The transport equations obtained in works [4, 7] are, in fact, the truncated sets of equations obtained in this paper; using such truncated equations may lead to losing accuracy in calculating flows and heat transfers in rarefied gases.

The integral representation for the 13 moments of the Boltzmann equation is presented in Section II, and the Hermite polynomial approximation of the velocity distribution function is derived in Section III. Grad's regularized 13 moment equations are obtained in Section IV, and conclusions are presented in Section V.

II.  GENERAL EQUATIONS FOR 13 MOMENTS

The phase density of a monatomic ideal gas is described by the Boltzmann equation,

$$\frac{\partial(n \cdot f)}{\partial t} + V_i \cdot \frac{\partial(n \cdot f)}{\partial x_i} = St(n \cdot f), \qquad \int_{\vec{V}} f \cdot d^3\vec{V} = 1, \qquad (1)$$

where $n$ is the number density of gas molecules, $f$ is the velocity distribution function, $V_i = (V_x, V_y, V_z)$ is the particle velocity, $x_i = (x, y, z)$ are the particle coordinates, the integral means the integration over the entire velocity space, and $St(n \cdot f)$ is the collision term that accounts for the change in the velocity distribution function due to collisions. Here we assume elastic collisions.

Let us introduce 13 moments of the particle distribution function: $\rho$ as the mass density of gas molecules, $u_i = (u_x, u_y, u_z)$ as the flow velocity of gas molecules, $V_T$ as the thermal velocity, $q_i = (q_x, q_y, q_z)$ as the heat flux, and $\sigma_{ij} = (\sigma_{xx}, \sigma_{xy}, \sigma_{xz}, \sigma_{yy}, \sigma_{zz})$ as the components of the stress tensor,

$$\rho = m \cdot n \cdot \int_{\vec{V}} f \cdot d^3\vec{V}, \quad u_i = \int_{\vec{V}} f \cdot V_i \cdot d^3\vec{V}, \quad \frac{3}{4} \cdot \rho \cdot V_T^2 = \frac{\rho}{2} \cdot \int_{\vec{V}} f \cdot (\vec{V} - \vec{u})^2 \cdot d^3\vec{V} \qquad (2)$$

$$q_i = \frac{\rho}{2} \cdot \int_{\vec{V}} f \cdot (V_i - u_i) \cdot (\vec{V} - \vec{u})^2 \cdot d^3\vec{V}, \quad \sigma_{ij} = \rho \cdot \int_{\vec{V}} f \cdot \left( (V_i - u_i) \cdot (V_j - u_j) - \delta_{ij} \cdot \frac{V_T^2}{2} \right) \cdot d^3\vec{V}, \qquad (3)$$

where $m$ is the mass of a particle and $(\vec{V} - \vec{u})^2 = (V_x - u_x)^2 + (V_y - u_y)^2 + (V_z - u_z)^2$. Multiplying Eq. (1) by

$$m, \quad m \cdot \vec{V}, \quad \frac{m}{2} \cdot \vec{V}^2, \quad \frac{m}{2} \cdot (\vec{V} - \vec{u}) \cdot (\vec{V} - \vec{u})^2, \quad m \cdot \left( (V_i - u_i) \cdot (V_j - u_j) - \delta_{ij} \cdot \frac{V_T^2}{2} \right) \quad (ij) = (xx, xy, xz, yy, yz), \qquad (4)$$

and then integrating the obtained equations over the entire velocity space, taking into account that the number of colliding particles, their total momentum, and their total energy are conserved in collision,

$$n \cdot \int_{\vec{V}} St(f) \cdot d^3\vec{V} = 0, \quad m \cdot n \cdot \int_{\vec{V}} V_i \cdot St(f) \cdot d^3\vec{V} = 0, \quad \frac{m}{2} \cdot n \cdot \int_{\vec{V}} \vec{V}^2 \cdot St(f) \cdot d^3\vec{V} = 0, \qquad (5)$$

after tedious algebra, the first five moment equations that correspond to mass, momentum and energy conservation laws can be derived to the form given in form [8],

$$\frac{\partial \rho}{\partial t} + \frac{\partial}{\partial x_i}(\rho \cdot u_i) = 0, \tag{6}$$

$$\rho \cdot \frac{\partial u_i}{\partial t} + (\rho \cdot u_j) \cdot \frac{\partial u_i}{\partial x_j} + \frac{\partial}{\partial x_i}\left(\frac{\rho \cdot V_T^2}{2}\right) + \frac{\partial \sigma_{ji}}{\partial x_j} = 0, \tag{7}$$

$$\frac{3}{4} \cdot \rho \cdot \frac{\partial V_T^2}{\partial t} + \frac{3}{2} \cdot \rho \cdot u_i \cdot \frac{\partial}{\partial x_i}\left(\frac{V_T^2}{2}\right) + \frac{1}{2} \cdot \rho \cdot V_T^2 \cdot \frac{\partial u_i}{\partial x_i} + \frac{\partial q_i}{\partial x_i} + \sigma_{ij} \cdot \frac{\partial u_i}{\partial x_j} = 0, \tag{8}$$

and the moment equations for $q_i$, $\sigma_{xy}$, and $\sigma_{xx}$ are

$$\frac{\partial q_i}{\partial t} + \frac{\partial}{\partial x_k}(u_k \cdot q_i) + q_k \cdot \frac{\partial u_i}{\partial x_k} - \left(\frac{5}{2} \cdot \frac{V_T^2}{2}\right) \cdot \left(\frac{\partial}{\partial x_i}\left(\frac{\rho \cdot V_T^2}{2} + \frac{\partial \sigma_{ij}}{\partial x_j}\right)\right) - \frac{\sigma_{ik}}{\rho} \cdot \left(\frac{\partial}{\partial x_k}\left(\frac{\rho \cdot V_T^2}{2}\right) + \frac{\partial \sigma_{jk}}{\partial x_j}\right) +$$
$$+ \frac{\partial}{\partial x_k}\left(\frac{\rho}{2} \cdot \int_{\vec{V}}(V_i - u_i) \cdot (\vec{V} - \vec{u})^2 \cdot (V_k - u_k) \cdot f \cdot d^3\vec{V}\right) + \left(\rho \cdot \int_{\vec{V}}(V_i - u_i) \cdot (V_k - u_k) \cdot (V_j - u_j) \cdot f \cdot d^3\vec{V}\right) \cdot \frac{\partial u_j}{\partial x_k} = \tag{9}$$
$$= \frac{\rho}{2} \cdot \int_{\vec{V}}(V_i - u_i) \cdot (\vec{V} - \vec{u})^2 \cdot St(f) \cdot d^3\vec{V},$$

$$\frac{\partial \sigma_{xy}}{\partial t} + \sigma_{ky} \cdot \frac{\partial u_x}{\partial x_k} + \sigma_{kx} \cdot \frac{\partial u_y}{\partial x_k} + \frac{\partial}{\partial x_k}(u_k \cdot \sigma_{xy}) + \rho \cdot V_T^2 \cdot \left(\frac{\partial u_x}{\partial y} + \frac{\partial u_y}{\partial x}\right) +$$
$$+ \frac{\partial}{\partial x_k}\left(\rho \cdot \int_{\vec{V}}(V_k - u_k) \cdot (V_x - u_x) \cdot (V_y - u_y) \cdot f \cdot d^3\vec{V}\right) = \rho \cdot \int_{\vec{V}}(V_x - u_x) \cdot (V_y - u_y) \cdot St(f) \cdot d^3\vec{V}, \tag{10}$$

$$\frac{\partial \sigma_{xx}}{\partial t} - \frac{1}{3} \cdot \rho \cdot V_T^2 \cdot \frac{\partial u_i}{\partial x_i} - \frac{2}{3} \cdot \frac{\partial q_i}{\partial x_i} - \frac{2}{3} \cdot \sigma_{ij} \cdot \frac{\partial u_i}{\partial x_j} + \frac{\partial}{\partial x_k}(u_k \cdot \sigma_{xx}) + 2 \cdot \sigma_{xk} \cdot \frac{\partial u_x}{\partial x_k} + \rho \cdot V_T^2 \cdot \frac{\partial u_x}{\partial x} +$$
$$+ \frac{\partial}{\partial x_k}\left(\rho \cdot \int_{\vec{V}}\left((V_x - u_x)^2 - \frac{V_T^2}{2}\right) \cdot (V_k - u_k) \cdot f \cdot d^3\vec{V}\right) = \rho \cdot \int_{\vec{V}}\left((V_x - u_x)^2 - \frac{V_T^2}{2}\right) \cdot St(f) \cdot d^3\vec{V}, \tag{11}$$

where $\sigma_{ij} = \sigma_{ji}$ and

$$\sigma_{zz} = \rho \cdot \int_{\vec{V}} f \cdot (V_z - u_z)^2 \cdot d^3\vec{V} - \frac{\rho \cdot V_T^2}{2} = -\sigma_{xx} - \sigma_{yy}. \tag{12}$$

Changing the order of indexes $(x, y, z)$ to $(x, z, y)$ and $(y, z, x)$ in Eq. (10), we obtain equations for $\sigma_{xz}$ and $\sigma_{yz}$ respectively, and to $(y, x, z)$ in Eq. (11), an equation for $\sigma_{yy}$. Derivation of Eq. (9) is presented in Appendix A; Eqs. (10) and (11) can be derived in a similar way.

It is worth noting that there are no collision terms in the moment equations for $\rho$, $u_i$, and $V_T^2$, Eqs. (6) – (8), while equations for heat flux and stress tensor, Eqs. (9) - (11), do include collision terms. Since collision terms can produce new moments, the set of these 13 moment equations in general is not self-contained. However, in the case of Maxwell molecules and the BGK approximation of the collision term,

$$St(f) = \frac{f_M - f}{\tau}, \qquad (13)$$

where $\tau$ is the collision time depending on coordinates and time, and $f_M$ is the Maxwellian velocity distribution function,

$$f_M = \left(\frac{1}{\pi \cdot V_T^2}\right)^{3/2} \cdot \exp\left(-\frac{(\vec{V} - \vec{u})^2}{V_T^2}\right), \qquad (14)$$

the collision terms do not produce any new moments. In other words, these approximations of the collision term *do not mix the moments*. This is a key point for any theory of moment approximation of the Boltzmann equation. In this paper we will use the BGK collision term; the case of Maxwellian molecules can be described in a similar way.

III. A HERMITE APPROXIMATION OF VELOCITY DISTRIBUTION FUNCTION

We assume here a Hermite polynomial approximation of the velocity distribution function where Hermite polynomials are described as follows:

$$H_N(\chi) = (-1)^N \cdot \exp(\chi^2) \cdot \frac{d^N}{d\chi^N}\left(\exp(-\chi^2)\right), \qquad (15)$$

$$\int_{-\infty}^{+\infty} H_N(\chi) \cdot H_M(\chi) \cdot \exp(-\chi^2) \cdot d\chi = \begin{cases} 0 & \text{if } N \neq M \\ \sqrt{\pi} \cdot N! \cdot 2^N & \text{if } N = M \end{cases}. \qquad (16)$$

The velocity distribution function can be described as a combination of three-dimensional Hermite polynomials that correspond to the $x$, $y$, $z$ directions of velocity. For our purposes, we need only the following set of Hermite polynomials:

$$H_0 = 1, \quad H_{1i}(\chi_i) = 2 \cdot \chi_i, \quad H_{2i}(\chi_i) = 4 \cdot \chi_i^2 - 2, \tag{17}$$

$$H_{3i}(\chi) = 8 \cdot \chi_i^3 - 12 \cdot \chi_i, \quad H_{4i}(\chi) = 16 \cdot \chi_i^4 - 48 \cdot \chi_i^2 + 12, \tag{18}$$

where $i = x, y, z$ and

$$\chi_i = \frac{V_i - u_i}{V_T}. \tag{19}$$

We represent the velocity distribution via the 35 Hermite polynomials

$$f_H = f_M \cdot \sum_{k=1}^{35} \Lambda_k \cdot \hat{H}_k\left(\chi_x, \chi_y, \chi_z\right), \tag{20}$$

where $f_M$ is a Maxwellian velocity distribution function, Eq. (14), coefficients $\Lambda$ depend on coordinates and time, and Hermite polynomials $\hat{H}$ are

$$\hat{H}_1 = H_0, \tag{21}$$

$$\hat{H}_2 = \frac{H_{1x}}{2}, \quad \hat{H}_3 = \frac{H_{1y}}{2}, \quad \hat{H}_4 = \frac{H_{1z}}{2}, \tag{22}$$

$$\hat{H}_5 = \frac{H_{2x}}{4}, \quad \hat{H}_6 = \frac{H_{1x} \cdot H_{1y}}{4}, \quad \hat{H}_7 = \frac{H_{1x} \cdot H_{1z}}{4}, \quad \hat{H}_8 = \frac{H_{2y}}{4}, \quad \hat{H}_9 = \frac{H_{1y} \cdot H_{1z}}{4}, \quad \hat{H}_{10} = \frac{H_{2z}}{4}, \tag{23}$$

$$\hat{H}_{11} = \frac{H_{3x} + H_{1x} \cdot H_{2y} + H_{1x} \cdot H_{2z}}{8}, \quad \hat{H}_{12} = \frac{H_{3y} + H_{1y} \cdot H_{2x} + H_{1y} \cdot H_{2z}}{8},$$

$$\hat{H}_{13} = \frac{H_{3z} + H_{1z} \cdot H_{2x} + H_{1z} \cdot H_{2y}}{8}, \tag{24}$$

and

$$\hat{H}_{14} = H_{4x}, \ \hat{H}_{15} = H_{4y}, \ \hat{H}_{16} = H_{4z}, \tag{25}$$

$$\hat{H}_{17} = H_{2x} \cdot H_{2y}, \ \hat{H}_{18} = H_{2x} \cdot H_{2z}, \ \hat{H}_{19} = H_{2y} \cdot H_{2z}, \tag{26}$$

$$\hat{H}_{20} = H_{3x} \cdot H_{1y}, \ \hat{H}_{21} = H_{3x} \cdot H_{1z}, \ \hat{H}_{22} = H_{3y} \cdot H_{1x}, \ \hat{H}_{23} = H_{3y} \cdot H_{1z},$$
$$\hat{H}_{24} = H_{3z} \cdot H_{1x}, \ \hat{H}_{25} = H_{3z} \cdot H_{1y}, \tag{27}$$

$$\hat{H}_{26} = H_{1x} \cdot H_{1y} \cdot H_{2z}, \ \hat{H}_{27} = H_{1y} \cdot H_{1z} \cdot H_{2x}, \ \hat{H}_{28} = H_{1z} \cdot H_{1x} \cdot H_{2y}, \tag{28}$$

$$\hat{H}_{29} = H_{1x} \cdot H_{1y} \cdot H_{1z}. \tag{29}$$

$$\hat{H}_{30} = H_{3x} - \frac{3}{2} \cdot H_{1x} \cdot (H_{2y} + H_{2z}), \ \hat{H}_{31} = H_{3y} - \frac{3}{2} \cdot H_{1y} \cdot (H_{2x} + H_{2z}),$$
$$\hat{H}_{32} = H_{3z} - \frac{3}{2} \cdot H_{1z} \cdot (H_{2x} + H_{2y}), \tag{30}$$

$$\hat{H}_{33} = H_{1x} \cdot (H_{2y} - H_{2z}), \ \hat{H}_{34} = H_{1y} \cdot (H_{2x} - H_{2z}), \ \hat{H}_{35} = H_{1z} \cdot (H_{2x} - H_{2y}). \tag{31}$$

It should be stressed that all $\hat{H}$ polynomials are orthogonal to each other, i.e.

$$\int_{-\infty}^{+\infty} \hat{H}_i \cdot \hat{H}_j \cdot \exp\left(-\chi_x^2 - \chi_y^2 - \chi_z^2\right) \cdot d^3\vec{\chi}, \text{ where } i \neq j. \tag{32}$$

Since the velocity distribution function $f_H$ has to satisfy Eqs. (2), we obtain that

$$\Lambda_1 = 1, \quad \Lambda_2 = \Lambda_3 = \Lambda_4 = 0 \quad \text{and} \quad \Lambda_5 + \Lambda_8 + \Lambda_{10} = 0. \tag{33}$$

Thus, the particle distribution function $n \cdot f_H$ has 35 variables, $\rho$, $u_x$, $u_y$, $u_z$, $V_T^2$, $\Lambda_5$-$\Lambda_9$, $\Lambda_{11}$-$\Lambda_{35}$. Substituting the velocity distribution function $f_H$, Eq. (20), for $f$ in Eqs. (3), we obtain relationships between $\sigma_{ij}$, $q_i$ and $\Lambda$:

$$\Lambda_5 = \frac{2 \cdot \sigma_{xx}}{\rho \cdot V_T^2}, \ \Lambda_6 = \frac{4 \cdot \sigma_{xy}}{\rho \cdot V_T^2}, \ \Lambda_7 = \frac{4 \cdot \sigma_{xz}}{\rho \cdot V_T^2}, \ \Lambda_8 = \frac{2 \cdot \sigma_{yy}}{\rho \cdot V_T^2}, \ \Lambda_9 = \frac{4 \cdot \sigma_{yz}}{\rho \cdot V_T^2}, \ \Lambda_{10} = -\frac{2 \cdot \sigma_{xx} + 2 \cdot \sigma_{yy}}{\rho \cdot V_T^2}, \tag{33}$$

$$\Lambda_{11} = \frac{8 \cdot q_x}{5 \cdot \rho \cdot V_T^3}, \ \Lambda_{12} = \frac{8 \cdot q_y}{5 \cdot \rho \cdot V_T^3}, \ \Lambda_{13} = \frac{8 \cdot q_z}{5 \cdot \rho \cdot V_T^3}. \tag{34}$$

It is worth noting that the truncated velocity distribution function $f_H$ that consists of the first ten nonzero Hermite polynomials has the form of the Chapman-Enskog and Grad's velocity distribution functions [1, 2, 8]. In the next section, it will be shown why we have selected this representation of the velocity distribution function.

## IV. A CLOSURE OF GRAD'S 13 MOMENT EQUATIONS

Let us rewrite Eqs. (9) – (11) for the case of the BGK collision term, Eq. (13):

$$\frac{\partial q_i}{\partial t} + \langle .... \rangle = -\frac{q_i}{\tau}, \ \frac{\partial \sigma_{xx}}{\partial t} + \langle .... \rangle = -\frac{\sigma_{xx}}{\tau}, \ \frac{\partial \sigma_{xy}}{\partial t} + \langle .... \rangle = -\frac{\sigma_{xy}}{\tau}, \tag{35}$$

where $\langle ... \rangle$ are the left hand side terms in Eqs. (9) – (11) positioned after corresponding time derivatives. To complete this system of equations, a velocity distribution function has to be chosen. Grad [1, 2] has suggested his velocity distribution function, which is the truncated velocity distribution function $f_H$, Eq. (20), with the first 10 non-zero terms shown in Eq. (21) – (34),

$$f_{GRAD} = f_M \cdot \left[ \hat{H}_0 + \frac{8 \cdot q_x}{5 \cdot \rho \cdot V_T^3} \cdot \hat{H}_{11} + \frac{8 \cdot q_y}{5 \cdot \rho \cdot V_T^3} \cdot \hat{H}_{12} + \frac{8 \cdot q_z}{5 \cdot \rho \cdot V_T^3} \cdot \hat{H}_{13} + \frac{2 \cdot \sigma_{xx}}{\rho \cdot V_T^2} \cdot \hat{H}_5 + \frac{4 \cdot \sigma_{xy}}{\rho \cdot V_T^2} \cdot \hat{H}_6 + \right.$$
$$\left. + \frac{4 \cdot \sigma_{xz}}{\rho \cdot V_T^2} \cdot \hat{H}_7 + \frac{2 \cdot \sigma_{yy}}{\rho \cdot V_T^2} \cdot \hat{H}_8 + \frac{4 \cdot \sigma_{yz}}{\rho \cdot V_T^2} \cdot \hat{H}_9 - \frac{2 \cdot (\sigma_{xx} + \sigma_{yy})}{\rho \cdot V_T^2} \cdot \hat{H}_{10} \right], \tag{36}$$

and obtained his set of 13 moment equations [1, 2], which can be presented in the following form:

$$\frac{\partial q_x}{\partial t} + \frac{\partial}{\partial x_k} \cdot (u_k \cdot q_x) + q_k \cdot \frac{\partial u_x}{\partial x_k} - \left(\frac{5 \cdot V_T^2}{8}\right) \cdot \frac{\partial}{\partial x}(\rho \cdot V_T^2) - \left(\frac{5 \cdot V_T^2}{4}\right) \cdot \frac{\partial \sigma_{jx}}{\partial x_j} - \left(\frac{\sigma_{xk}}{2 \cdot \rho}\right) \cdot \frac{\partial}{\partial x_k}(\rho \cdot V_T^2) - \frac{\sigma_{xk}}{\rho} \cdot \frac{\partial \sigma_{jk}}{\partial x_j} +$$
$$+ \frac{\partial}{\partial x}\left(\frac{5 \cdot \rho \cdot V_T^4}{8} + \frac{7 \cdot V_T^2 \cdot \sigma_{xx}}{4}\right) + \frac{\partial}{\partial y}\left(\frac{7 \cdot V_T^2 \cdot \sigma_{xy}}{4}\right) + \frac{\partial}{\partial z}\left(\frac{7 \cdot V_T^2 \cdot \sigma_{xz}}{4}\right) + \left(\frac{6 \cdot q_x}{5}\right) \cdot \frac{\partial u_x}{\partial x} + \left(\frac{2 \cdot q_y}{5}\right) \cdot \frac{\partial u_y}{\partial x} + \quad (37)$$
$$+ \left(\frac{2 \cdot q_y}{5}\right) \cdot \frac{\partial u_x}{\partial y} + \left(\frac{2 \cdot q_z}{5}\right) \cdot \frac{\partial u_z}{\partial x} + \left(\frac{2 \cdot q_z}{5}\right) \cdot \frac{\partial u_x}{\partial z} + \left(\frac{2 \cdot q_x}{5}\right) \cdot \frac{\partial u_y}{\partial y} + \left(\frac{2 \cdot q_x}{5}\right) \cdot \frac{\partial u_z}{\partial z} = -\frac{q_x}{\tau},$$

$$\frac{\partial \sigma_{xx}}{\partial t} + \frac{\partial}{\partial x_k}(u_k \cdot \sigma_{xx}) + (2 \cdot \sigma_{xx}) \cdot \frac{\partial u_x}{\partial x} + (2 \cdot \sigma_{xy}) \cdot \frac{\partial u_x}{\partial y} + (2 \cdot \sigma_{xz}) \cdot \frac{\partial u_x}{\partial z} - \left(\frac{2 \cdot \sigma_{kl}}{3}\right) \cdot \frac{\partial u_l}{\partial x_k} +$$
$$+ \left(\frac{2 \cdot \rho \cdot V_T^2}{3}\right) \cdot \frac{\partial u_x}{\partial x} - \left(\frac{\rho \cdot V_T^2}{3}\right) \cdot \frac{\partial u_y}{\partial y} - \left(\frac{\rho \cdot V_T^2}{3}\right) \cdot \frac{\partial u_z}{\partial z} + \frac{8}{15} \cdot \frac{\partial q_x}{\partial x} - \frac{4}{15} \cdot \frac{\partial q_y}{\partial y} - \frac{4}{15} \cdot \frac{\partial q_z}{\partial z} = -\frac{\sigma_{xx}}{\tau}, \quad (38)$$

$$\frac{\partial \sigma_{xy}}{\partial t} + \frac{\partial}{\partial x_k}(u_k \cdot \sigma_{xy}) + \sigma_{ky} \cdot \frac{\partial u_x}{\partial x_k} + \left(\frac{\rho \cdot V_T^2}{2}\right) \cdot \frac{\partial u_x}{\partial y} + \sigma_{kx} \cdot \frac{\partial u_y}{\partial x_k} + \left(\frac{\rho \cdot V_T^2}{2}\right) \cdot \frac{\partial u_y}{\partial x} + \frac{2}{5} \cdot \frac{\partial q_y}{\partial x} + \frac{2}{5} \cdot \frac{\partial q_x}{\partial y} = -\frac{\sigma_{xy}}{\tau}. \quad (39)$$

Changing the order of indexes $(x, y, z)$ to $(y, x, z)$ and $(z, y, x)$ in Eq. (37), we obtain equations for $q_x$ and $q_z$ respectively; likewise, using $(y, x, z)$ in Eq. (38), we obtain an equation for $\sigma_{yy}$; and finally, using $(x, z, y)$ and $(y, z, x)$ in Eq. (39) we obtain equations for $\sigma_{xz}$ and $\sigma_{yz}$ respectively. Unfortunately, as it has been mentioned in the introduction, Grad's moment equations sometimes produce unphysical solutions and, therefore, need to be regularized.

Let us obtain a set of the 13 regularized Grad's moment equations using the Hermite polynomial approximation of the velocity distribution function and the Chapman-Enskog closure method, while assuming that the collision term is in BGK form, Eq. (14). First, let us represent the under integral polynomials in the left-hand sides of Eqs. (9) for $q_x$ and Eqs. (10) and (11) in Hermite forms:

$$\chi_x^2 \cdot (\chi_x^2 + \chi_y^2 + \chi_z^2) = \frac{H_{4x}}{16} + \frac{H_{2x} \cdot H_{2y}}{4} + \frac{H_{2x} \cdot H_{2z}}{4} + \frac{7 \cdot H_{2x}}{4} + \frac{H_{2z}}{2} + \frac{H_{2y}}{2} + \frac{5 \cdot H_0}{4}, \quad (40)$$

$$\chi_x \cdot \chi_y \cdot (\chi_x^2 + \chi_y^2 + \chi_z^2) = \frac{H_{3x} \cdot H_{1y}}{16} + \frac{H_{3y} \cdot H_{1x}}{16} + \frac{H_{1x} \cdot H_{1y} \cdot H_{2z}}{16} + \frac{7 \cdot H_{1x} \cdot H_{1y}}{8}, \quad (41)$$

$$\chi_x \cdot \chi_z \cdot \left(\chi_x^2 + \chi_y^2 + \chi_z^2\right) = \frac{H_{3x} \cdot H_{1z}}{16} + \frac{H_{3z} \cdot H_{1x}}{16} + \frac{H_{1x} \cdot H_{1z} \cdot H_{2y}}{16} + \frac{7 \cdot H_{1x} \cdot H_{1z}}{8}, \qquad (42)$$

$$\chi_x^3 = \frac{H_{3x}}{8} + \frac{3 \cdot H_{1x}}{2}, \qquad (43)$$

$$\chi_x^2 \cdot \chi_y = \frac{H_{2x} \cdot H_{1y}}{8} + \frac{H_{1y}}{4}, \qquad (44)$$

$$\chi_x^2 \cdot \chi_z = \frac{H_{2x} \cdot H_{1z}}{8} + \frac{H_{1z}}{4}, \qquad (45)$$

$$\chi_x \cdot \chi_y \cdot \chi_z = \frac{H_{1x} \cdot H_{1y} \cdot H_{1z}}{8}, \qquad (46)$$

$$\chi_x \cdot \chi_y^2 = \frac{H_{2y} \cdot H_{1x}}{8} + \frac{H_{1x}}{4}, \qquad (47)$$

$$\chi_x \cdot \chi_z^2 = \frac{H_{2z} \cdot H_{1x}}{8} + \frac{H_{1x}}{4}, \qquad (48)$$

$$\left(\chi_x^2 - \frac{1}{2}\right) \cdot \chi_y = \frac{H_{2x} \cdot H_{1y}}{8}, \qquad (49)$$

$$\left(\chi_x^2 - \frac{1}{2}\right) \cdot \chi_x = \frac{H_{3x}}{8} + \frac{H_{1x}}{2}, \qquad (50)$$

$$\left(\chi_x^2 - \frac{1}{2}\right) \cdot \chi_z = \frac{H_{2x} \cdot H_{1z}}{8}, \qquad (51)$$

where variables $\chi$ are given by Eq. (19). As one can see, the polynomials presented in: Eqs. (40) – (48) are included in $q_x$; Eqs. (44), (46) and (47) in $\sigma_{xy}$; and Eqs. (49) – (51) in $\sigma_{xx}$. Thus, the complete list of Hermite polynomials that represents the under integral polynomials in equations for heat flux and stress tensor is

$$H_{4x},\ H_{4y},\ H_{4z}, \tag{52}$$

$$H_{2x}\cdot H_{2y},\ H_{2x}\cdot H_{2z},\ H_{2y}\cdot H_{2z}, \tag{53}$$

$$H_{3x}\cdot H_{1y},\ H_{3x}\cdot H_{1z},\ H_{3y}\cdot H_{1x},\ H_{3y}\cdot H_{1z},\ H_{3z}\cdot H_{1x},\ H_{3z}\cdot H_{1y}, \tag{54}$$

$$H_{1x}\cdot H_{1y}\cdot H_{2z},\ H_{1x}\cdot H_{1z}\cdot H_{2y},\ H_{1z}\cdot H_{1y}\cdot H_{2x}, \tag{55}$$

$$H_{1x}\cdot H_{1y}\cdot H_{1z}, \tag{56}$$

$$H_{3x},\ H_{3y},\ H_{3z}, \tag{57}$$

$$H_{2x}\cdot H_{1y},\ H_{2x}\cdot H_{1z},\ H_{2y}\cdot H_{1x},\ H_{2y}\cdot H_{1z},\ H_{2z}\cdot H_{1x},\ H_{2z}\cdot H_{1y}, \tag{58}$$

$$H_{2x},\ H_{2y},\ H_{2z}, \tag{59}$$

$$H_{1x}\cdot H_{1y},\ H_{1x}\cdot H_{1z},\ H_{1y}\cdot H_{1z}, \tag{60}$$

$$H_{1x},\ H_{1y},\ H_{1z}, \tag{61}$$

$$H_0. \tag{62}$$

The total number of these polynomials is 35, and all of them are orthogonal to each other. As one can see the Hermite polynomials in Eqs. (21) – (31) are identical to the polynomials in Eqs. (52) – (62), or are their linear combinations, Eqs. (24), (30), (31) and, therefore, also can be used to represent the under integral polynomials in equations for heat flux and stress tensor. Thus, we have shown that the chosen set of Hermite polynomials, Eq. (21) – (33), is complete and has good physical sense for representing the velocity distribution function.

Now, following the Chapman-Enskog method [5, 6], let us write the velocity distribution function as

$$f = f_{GRAD} + \tau \cdot f_M \cdot f_1, \tag{63}$$

where $f_1$ has a Hermite form. Since the velocity distribution function $\tau \cdot f_M \cdot f_1$ does not have to contribute to the previously obtained 13 moments $\rho$, $u_i$, $V_T^2$, $q_i$, and $\sigma_{ij}$, it follows that the Hermite polynomials included in Grad's velocity distribution function, Eqs. (21) – (24), have to be excluded from $f_1$. But as the velocity distribution function $\tau \cdot f_M \cdot f_1$ has to contribute to the integrals in Eqs. (9) – (11), we obtain that $f_1$ has to be a combination only of the Hermite polynomials presented in Eqs. (25) – (31). Subsequently, we obtain that

$$f_1 = \sum_{k=14}^{35} \Lambda_i \cdot \hat{H}_k, \tag{64}$$

Substituting $f_H$, Eq. (20), for $f$ into Eqs. (35), and introducing the following sets of $M$-moments,

$$M_{4i} = V_T^4 \cdot \rho \cdot \int_{\vec{V}} f_H \cdot H_{4i} \cdot d^3\vec{V}, \quad i = (x, y, z), \tag{65}$$

$$M_{2i2j} = V_T^4 \cdot \rho \cdot \int_{\vec{V}} f_H \cdot H_{2i} \cdot H_{2j} \cdot d^3\vec{V}, \quad ij = (xy, xz, yz), \tag{66}$$

$$M_{3i1j} = V_T^4 \cdot \rho \cdot \int_{\vec{V}} f_H \cdot H_{3i} \cdot H_{1j} \cdot d^3\vec{V}, \quad ij = (xy, xz, yx, yz, zx, zy), \tag{67}$$

$$M_{1i1j2k} = V_T^4 \cdot \rho \cdot \int_{\vec{V}} f_H \cdot H_{1i} \cdot H_{1j} \cdot H_{2k} \cdot d^3\vec{V}, \quad ijk = (xyz, yzx, zxy), \tag{68}$$

$$M_{1i1j1k} = V_T^3 \cdot \rho \cdot \int_{\vec{V}} f_H \cdot H_{1i} \cdot H_{1j} \cdot H_{1k} \cdot d^3\vec{V}, \quad ijk = (xyz), \tag{69}$$

$$M_{i(2j+2k)} = V_T^3 \cdot \rho \cdot \int_{\vec{V}} f_H \cdot \left( H_{3i} - \frac{3}{2} \cdot H_{1i} \cdot (H_{2j} + H_{2k}) \right) \cdot d^3\vec{V}, \quad ijk = (xyz, yxz, zxy), \tag{70}$$

$$M_{1i(2j-2k)} = V_T^3 \cdot \rho \cdot \int_{\vec{V}} f_H \cdot H_{1i} \cdot (H_{2j} - H_{2k}) \cdot d^3\vec{V}, \quad ijk = (xyz, yxz, zxy), \tag{71}$$

which correspond to Hermite polynomials in Eqs. (25) – (31), we obtain the following equations for $q_x$, $\sigma_{xx}$, and $\sigma_{xy}$:

$$\frac{\partial q_x}{\partial t} + \{....\} + \frac{\partial}{\partial x}\left(\frac{M_{4x}}{32} + \frac{M_{2x2y}}{8} + \frac{M_{2x2z}}{8}\right) + \frac{\partial}{\partial y}\left(\frac{M_{3x1y}}{32} + \frac{M_{1x3y}}{32} + \frac{M_{1x1y2z}}{32}\right) +$$
$$+ \frac{\partial}{\partial z}\left(\frac{M_{3x1z}}{32} + \frac{M_{1x3z}}{32} + \frac{M_{1x1z2y}}{32}\right) + \left(\frac{M_{x(2y+2z)}}{20}\right)\cdot\frac{\partial u_x}{\partial x} + \left(\frac{M_{1y(2x-2z)}}{16} - \frac{M_{y(2x+2z)}}{40}\right)\cdot\left(\frac{\partial u_x}{\partial y} + \frac{\partial u_y}{\partial x}\right) +$$
$$+ \left(\frac{M_{1z(2x-2y)}}{16} - \frac{M_{z(2x+2y)}}{40}\right)\cdot\left(\frac{\partial u_x}{\partial z} + \frac{\partial u_z}{\partial x}\right) + \frac{M_{1x1y1z}}{8}\cdot\left(\frac{\partial u_y}{\partial z} + \frac{\partial u_y}{\partial z}\right) = -\frac{q_x}{\tau}, \quad (72)$$

$$\frac{\partial \sigma_{xx}}{\partial t} + \{....\} + \frac{\partial}{\partial x}\left(\frac{M_{x(2y+2z)}}{20}\right) + \frac{\partial}{\partial y}\left(\frac{M_{1y(2x-2z)}}{16} - \frac{M_{y(2x+2z)}}{40}\right) +$$
$$+ \frac{\partial}{\partial z}\left(\frac{M_{1z(2x-2y)}}{16} - \frac{M_{z(2x+2y)}}{40}\right) = -\frac{\sigma_{xx}}{\tau}, \quad (73)$$

$$\frac{\partial \sigma_{xy}}{\partial t} + \{....\} + \frac{\partial}{\partial x}\left(\frac{M_{1y(2x-2z)}}{16} - \frac{M_{y(2x+2z)}}{40}\right) + \frac{\partial}{\partial y}\left(\frac{M_{1x(2y-2z)}}{16} - \frac{M_{x(2y+2z)}}{40}\right) +$$
$$+ \frac{1}{8}\cdot\frac{\partial}{\partial z}\left(M_{1x1y1z}\right) = -\frac{\sigma_{xy}}{\tau}, \quad (74)$$

where $\{....\}$ are terms due to Grad's part of the velocity distribution function $f$, Eq. (63); they are shown in Eqs. (37) – (39). The equations for $q_y$, $q_z$, $\sigma_{yy}$, $\sigma_{xz}$, and $\sigma_{yz}$ can be obtained by proper rotations of indexes in Eqs. (72) – (74). Thus, we have shown that the set of moments presented in Eqs. (65) – (71) is a complete set of moments needed for extension (regularization) of Grad's thirteen equations to the third order of the Knudsen number. Applying the Chapman-Enskog technique [5, 6] to Grad's velocity distribution function, Eq. (36), we obtain equations for the *M*-moments,

$$-M_{1x1y1z} = -\tau\cdot\left\{4\cdot\rho\cdot V_T^2\cdot\left[\frac{\partial}{\partial x}\left(\frac{\sigma_{yz}}{\rho}\right) + \frac{\partial}{\partial y}\left(\frac{\sigma_{xz}}{\rho}\right) + \frac{\partial}{\partial z}\left(\frac{\sigma_{xy}}{\rho}\right)\right] + \left(\frac{16\cdot q_z}{5}\right)\cdot\left(\frac{\partial u_x}{\partial y} + \frac{\partial u_y}{\partial x}\right) +$$
$$+ \left(\frac{16\cdot q_y}{5}\right)\cdot\left(\frac{\partial u_x}{\partial z} + \frac{\partial u_z}{\partial x}\right) + \left(\frac{16\cdot q_x}{5}\right)\cdot\left(\frac{\partial u_y}{\partial z} + \frac{\partial u_z}{\partial y}\right)\right\}. \quad (75)$$

$$-\frac{M_{2x2y}}{\tau} = -\frac{32 \cdot V_T^2}{5 \cdot \rho} \cdot \left( q_x \cdot \frac{\partial \rho}{\partial x} + q_y \cdot \frac{\partial \rho}{\partial x} \right) + \left( \frac{32 \cdot V_T^2}{5} \right) \cdot \left( \frac{\partial q_x}{\partial x} + \frac{\partial q_y}{\partial y} \right) +$$

$$+ \left( \frac{64 \cdot q_x}{5} \right) \cdot \frac{\partial V_T^2}{\partial x} + \left( \frac{64 \cdot q_y}{5} \right) \cdot \frac{\partial V_T^2}{\partial y} + \left( \frac{32 \cdot q_z}{5} \right) \cdot \frac{\partial V_T^2}{\partial z} + \left( 32 \cdot V_T^2 \cdot \sigma_{xy} \right) \cdot \left( \frac{\partial u_x}{\partial y} + \frac{\partial u_y}{\partial x} \right) + \quad (76)$$

$$+ \frac{16 \cdot V_T^2}{3} \cdot \left( (2 \cdot \sigma_{yy} - \sigma_{xx}) \cdot \frac{\partial u_x}{\partial x} + (2 \cdot \sigma_{xx} - \sigma_{yy}) \cdot \frac{\partial u_y}{\partial y} - (\sigma_{xx} + \sigma_{yy}) \cdot \frac{\partial u_z}{\partial z} \right),$$

$$-\frac{M_{4x}}{\tau} = \left( \frac{288}{5} \cdot q_x \right) \cdot \frac{\partial V_T^2}{\partial x} + \left( \frac{96}{5} \cdot q_y \right) \cdot \frac{\partial V_T^2}{\partial y} + \left( \frac{96}{5} \cdot q_z \right) \cdot \frac{\partial V_T^2}{\partial z} + \left( \frac{192}{5} \cdot V_T^2 \right) \cdot \frac{\partial q_x}{\partial x} -$$

$$- \left( \frac{192}{5} \cdot \frac{q_x \cdot V_T^2}{\rho} \right) \cdot \frac{\partial \rho}{\partial x} + \left( 32 \cdot \sigma_{xx} \cdot V_T^2 \right) \cdot \left( 2 \cdot \frac{\partial u_x}{\partial x} - \frac{\partial u_y}{\partial y} - \frac{\partial u_z}{\partial z} \right). \quad (77)$$

$$-\frac{M_{3x1y}}{\tau} = -16 \cdot \sigma_{xy} \cdot V_T^2 \cdot \left( \frac{\partial u_y}{\partial y} + \frac{\partial u_z}{\partial z} - 2 \cdot \frac{\partial u_x}{\partial x} \right) + 24 \cdot \sigma_{xx} \cdot V_T^2 \cdot \left( \frac{\partial u_x}{\partial y} + \frac{\partial u_y}{\partial x} \right) -$$

$$- \frac{48 \cdot V_T^2}{5 \cdot \rho} \cdot \left( q_y \cdot \frac{\partial \rho}{\partial x} + q_x \cdot \frac{\partial \rho}{\partial y} \right) + \frac{48 \cdot V_T^2}{5} \cdot \left( \frac{\partial q_y}{\partial x} + \frac{\partial q_x}{\partial y} \right) + \frac{48}{5} \cdot \left( q_y \cdot \frac{\partial V_T^2}{\partial x} + q_x \cdot \frac{\partial V_T^2}{\partial y} \right). \quad (78)$$

$$-\frac{M_{1x1y2z}}{\tau} = -\left( \frac{16}{3} \cdot \sigma_{xy} \cdot V_T^2 \right) \cdot \frac{\partial u_x}{\partial x} - \left( \frac{16}{3} \cdot \sigma_{xy} \cdot V_T^2 \right) \cdot \frac{\partial u_y}{\partial y} + \left( 8 \cdot \sigma_{zz} \cdot V_T^2 \right) \cdot \frac{\partial u_x}{\partial y} + \left( 16 \cdot \sigma_{yz} \cdot V_T^2 \right) \cdot \frac{\partial u_x}{\partial z} + \left( 8 \cdot \sigma_{zz} \cdot V_T^2 \right) \cdot \frac{\partial u_y}{\partial x} +$$

$$+ \left( 16 \cdot \sigma_{xz} \cdot V_T^2 \right) \cdot \frac{\partial u_y}{\partial z} + \left( 16 \cdot \sigma_{yz} \cdot V_T^2 \right) \cdot \frac{\partial u_z}{\partial x} + \left( 16 \cdot \sigma_{xz} \cdot V_T^2 \right) \cdot \frac{\partial u_z}{\partial y} + \left( \frac{32}{3} \cdot \sigma_{xy} \cdot V_T^2 \right) \cdot \frac{\partial u_z}{\partial z} - \left( \frac{16}{5} \cdot \frac{q_y \cdot V_T^2}{\rho} \right) \cdot \frac{\partial \rho}{\partial x} -$$

$$- \left( \frac{16}{5} \cdot \frac{q_x \cdot V_T^2}{\rho} \right) \cdot \frac{\partial \rho}{\partial y} + \left( \frac{16}{5} \cdot V_T^2 \right) \cdot \frac{\partial q_y}{\partial x} + \left( \frac{16}{5} \cdot V_T^2 \right) \cdot \frac{\partial q_x}{\partial y} + \left( \frac{16}{5} \cdot q_y \right) \cdot \frac{\partial V_T^2}{\partial x} + \left( \frac{16}{5} \cdot q_x \right) \cdot \frac{\partial V_T^2}{\partial y}.$$

$$(79)$$

$$-\frac{M_{x(2y+2z)}}{\tau} = -6 \cdot V_T^2 \cdot \left( 3 \cdot \sigma_{xx} \cdot \frac{d(\ln \rho)}{dx} - \sigma_{xy} \cdot \frac{d(\ln \rho)}{dy} - \sigma_{xz} \cdot \frac{d(\ln \rho)}{dz} \right) + 6 \cdot V_T^2 \cdot \left( 3 \cdot \frac{d\sigma_{xx}}{dx} - 2 \cdot \frac{d\sigma_{xy}}{dy} - 2 \cdot \frac{d\sigma_{xz}}{dz} \right) +$$

$$+ \frac{48 \cdot q_x}{5} \cdot \left( 2 \cdot \frac{du_x}{dx} - \frac{du_y}{dy} - \frac{du_z}{dz} \right) - \frac{48 \cdot q_y}{5} \cdot \left( \frac{du_y}{dx} + \frac{du_x}{dy} \right) - \frac{48 \cdot q_z}{5} \cdot \left( \frac{du_z}{dx} + \frac{du_x}{dz} \right),$$

$$(80)$$

$$-\frac{M_{1x(2y-2z)}}{\tau} = 4 \cdot V_T^2 \cdot \left( \frac{d\sigma_{yy}}{dx} + 2 \cdot \frac{d\sigma_{xy}}{dy} - \frac{d\sigma_{zz}}{dx} - 2 \cdot \frac{d\sigma_{xz}}{dz} \right) +$$
$$+ \frac{32 \cdot q_x}{5} \cdot \left( \frac{du_y}{dy} - \frac{du_z}{dz} \right) + \frac{32 \cdot q_y}{5} \cdot \left( \frac{du_y}{dx} + \frac{du_x}{dy} \right) - \frac{32 \cdot q_z}{5} \cdot \left( \frac{du_z}{dx} + \frac{du_x}{dz} \right) + \quad (81)$$
$$+ 4 \cdot V_T^2 \cdot \left( 2 \cdot \sigma_{xz} \cdot \frac{d(\ln \rho)}{dz} - 2 \cdot \sigma_{xy} \cdot \frac{d(\ln \rho)}{dy} + (\sigma_{zz} - \sigma_{yy}) \cdot \frac{d(\ln \rho)}{dz} \right).$$

Derivation of the equation for $M_{1x1y1z}$ is presented in Appendix B; other $M$-moments can be obtained in a similar way. Changing the order of indexes $(x, y, z)$ to $(x, z, y)$ in Eq. (76), we obtain an equation for the $M_{2x2z}$ moment, similarly for the $M_{2y2z}$ moment; changing the order of indexes $(x, y, z)$ to $(y, x, z)$ in Eq. (77), we obtain an equation for the $M_{4y}$ moment, similarly, for the $M_{4z}$ moment; changing the order of indexes $(x, y, z)$ to $(x, z, y)$ and $(y, x, z)$ in Eq. (78), we obtain equations for the $M_{3x1z}$ and $M_{3y1x}$ moments, respectively, and similarly for the $M_{3y1z}$, $M_{3z1y}$, and $M_{3z1x}$ moments. Changing the order of indexes $(x, y, z)$ to $(x, z, y)$ in Eq. (79), we obtain an equation for the $M_{1x1z3y}$ moment, and using $(z, y, x)$ we obtain an equation for the $M_{1y1z3x}$ moment. Changing the order of indexes $(x, y, z)$ to $(y, x, z)$ and $(z, x, y)$ in Eq. (80), we obtain equations for $M_{y(2x+2z)}$ and $M_{z(2x+2y)}$, respectively. Changing the order of indexes $(x, y, z)$ to $(y, x, z)$ and $(z, x, y)$ in Eq. (81), we obtain equations for $M_{1y(2x-2z)}$ and $M_{1z(2x-2y)}$, respectively.

Thus, the resulting set of 13 equations, Eqs. (6) – (8), (72) – (74), with $M$-moments from Eqs. (78) – (81) with rotations of indexes as described above, represents a complete set of the 13 regularized Grad's moment equations for $\rho$, $(u_x, u_y, u_z)$, $V_T^2$, $(q_x, q_y, q_z)$, and $(\sigma_{xx}, \sigma_{xy}, \sigma_{xz}, \sigma_{yy}, \sigma_{yz})$.

Finally, substituting the velocity distribution function $f_H$, Eq. (20), in Eqs. (65) – (71) we obtain coefficients $\Lambda_{14}$ - $\Lambda_{35}$,

$$\Lambda_{14} = \frac{M_{4x}}{384 \cdot \rho \cdot V_T^4}, \quad \Lambda_{15} = \frac{M_{4y}}{384 \cdot \rho \cdot V_T^4}, \quad \Lambda_{16} = \frac{M_{4z}}{384 \cdot \rho \cdot V_T^4}, \quad (82)$$

$$\Lambda_{17} = \frac{M_{2x2y}}{64 \cdot \rho \cdot V_T^4}, \quad \Lambda_{18} = \frac{M_{2x2z}}{64 \cdot \rho \cdot V_T^4}, \quad \Lambda_{19} = \frac{M_{2x2z}}{64 \cdot \rho \cdot V_T^4}, \quad (83)$$

$$\Lambda_{20} = \frac{M_{3x1y}}{96 \cdot \rho \cdot V_T^4} \,, \quad \Lambda_{21} = \frac{M_{3x1z}}{96 \cdot \rho \cdot V_T^4} \,, \quad \Lambda_{22} = \frac{M_{3y1x}}{96 \cdot \rho \cdot V_T^4} \,, \quad \Lambda_{23} = \frac{M_{3y1z}}{96 \cdot \rho \cdot V_T^4} \,,$$

$$\Lambda_{24} = \frac{M_{3z1x}}{96 \cdot \rho \cdot V_T^4} \,, \quad \Lambda_{25} = \frac{M_{3z1y}}{96 \cdot \rho \cdot V_T^4} \,, \tag{84}$$

$$\Lambda_{26} = \frac{M_{1x1y2z}}{32 \cdot \rho \cdot V_T^4} \,, \quad \Lambda_{27} = \frac{M_{1y1z2x}}{32 \cdot \rho \cdot V_T^4} \,, \quad \Lambda_{28} = \frac{M_{1x1z2y}}{32 \cdot \rho \cdot V_T^4} \,, \tag{85}$$

$$\Lambda_{29} = \frac{M_{1x1y1z}}{8 \cdot \rho \cdot V_T^3} \,, \tag{86}$$

$$\Lambda_{30} = \frac{M_{x(2y+2z)}}{120 \cdot \rho \cdot V_T^3} \,, \quad \Lambda_{31} = \frac{M_{y(2x+2z)}}{120 \cdot \rho \cdot V_T^3} \,, \quad \Lambda_{32} = \frac{M_{z(2x+2y)}}{120 \cdot \rho \cdot V_T^3} \,, \tag{87}$$

$$\Lambda_{33} = \frac{M_{1x(2y-2z)}}{32 \cdot \rho \cdot V_T^3} \,, \quad \Lambda_{34} = \frac{M_{1y(2x-2z)}}{32 \cdot \rho \cdot V_T^3} \,, \quad \Lambda_{35} = \frac{M_{1z(2x-2y)}}{32 \cdot \rho \cdot V_T^3} \,. \tag{88}$$

## V. CONCLUSIONS

We have presented a new set of moment equations for rarefied gas dynamics. Our equations are a closure for Grad's 13 moment equations extended to the third order of the Knudsen number. We have assumed a Hermite polynomial approximation for the monatomic gas velocity distribution function and the BGK approximation of the collision term in the Boltzmann kinetic equation. We have also used the well-known Chapman-Enskog regularization method [5, 6], which has previously been used to derive a closure of Euler's gas dynamic equations. We have shown that the selected 35-term Hermite polynomial representation of the velocity distribution function is a complete set of polynomials for the closure of Grad's 13 moment equations with the third order of the Knudsen layer. Using a truncated set of basis functions may lead to a loss of accuracy in computing flows and heat transfer in rarefied gases. Our regularized Grad's 13 moment equations differ from the set of equations obtained in [4], where the authors use 26 polynomials for representation of the velocity distribution function, and a very different and complicated method. On the contrary, the closure method presented in this paper turns out to be quite straightforward and comprehensible. It can been shown that the gas dynamics equations [4] can be obtained by our method, and are a truncated set of our equations. Closure for Grad's 13 moment equations [7], where we use the 29 Hermite polynomial representation of the gas velocity distribution function, is also a truncated set of these equations.

APPENDIX A

Let us obtain the general moment equation for $q_i$. Multiplying the Boltzmann Equation, Eq. (1), by

$$\frac{m}{2} \cdot (V_i - u_i) \cdot \left[ (V_x - u_x)^2 + (V_y - u_y)^2 + (V_z - u_z)^2 \right] , \tag{A1}$$

we obtain

$$\frac{m}{2} \cdot (V_i - u_i) \cdot \left[ (V_x - u_x)^2 + (V_y - u_y)^2 + (V_z - u_z)^2 \right] \cdot \frac{\partial (n \cdot f)}{\partial t} +$$
$$+ \frac{m}{2} \cdot (V_i - u_i) \cdot V_k \cdot \left[ (V_x - u_x)^2 + (V_y - u_y)^2 + (V_z - u_z)^2 \right] \cdot \frac{\partial (f \cdot n)}{\partial x_k} = \tag{A2}$$
$$= \frac{m}{2} \cdot (V_i - u_i) \cdot \left[ (V_x - u_x)^2 + (V_y - u_y)^2 + (V_z - u_z)^2 \right] \cdot St(n \cdot f) .$$

Transferring the terms in front of the derivatives inside the derivative brackets in Eq. (A2), we obtain

$$\frac{\partial}{\partial t}\left(\frac{\rho}{2} \cdot (V_i - u_i) \cdot (\vec{V} - \vec{u})^2 \cdot f\right) + \frac{\partial}{\partial t}\left(\frac{\rho}{2} \cdot (\vec{V} - \vec{u})^2 \cdot f\right) \cdot \frac{\partial u_i}{\partial t} + \left(\rho \cdot (V_i - u_i) \cdot (V_k - u_k) \cdot f\right) \cdot \frac{\partial u_k}{\partial t} +$$
$$+ \frac{\partial}{\partial x_k}\left(\frac{\rho}{2} \cdot (V_i - u_i) \cdot (V_k - u_k) \cdot (\vec{V} - \vec{u})^2 \cdot f\right) + \frac{\partial}{\partial x_k}\left(u_k \cdot \frac{\rho}{2} \cdot (V_i - u_i) \cdot (\vec{V} - \vec{u})^2 \cdot f\right) +$$
$$+ \left(\frac{\rho}{2} \cdot (V_k - u_k) \cdot (\vec{V} - \vec{u})^2 \cdot f\right) \cdot \frac{\partial u_i}{\partial x_k} + u_k \cdot \left(\frac{\rho}{2} \cdot (\vec{V} - \vec{u})^2 \cdot f\right) \cdot \frac{\partial u_i}{\partial x_k} + \tag{A3}$$
$$+ \left(\rho \cdot (V_i - u_i) \cdot (V_k - u_k) \cdot (V_j - u_j) \cdot f\right) \cdot \frac{\partial u_j}{\partial x_k} + u_k \cdot \left(\rho \cdot (V_i - u_i) \cdot (V_j - u_j) \cdot f\right) \cdot \frac{\partial u_j}{\partial x_k} =$$
$$= \frac{\rho}{2} \cdot (V_i - u_i) \cdot (\vec{V} - \vec{u})^2 \cdot St(f) .$$

Integrating Eq. (A3) over the entire velocity domain, we obtain

$$\frac{\partial}{\partial t}\left(\frac{\rho}{2}\cdot\int_{\vec{V}}(V_i-u_i)\cdot(\vec{V}-\vec{u})^2\cdot f\cdot d^3\vec{V}\right)+\frac{\partial}{\partial t}\left(\frac{\rho}{2}\cdot\int_{\vec{V}}(\vec{V}-\vec{u})^2\cdot f\cdot d^3\vec{V}\right)\cdot\frac{\partial u_i}{\partial t}+$$

$$+\left(\rho\cdot\int_{\vec{V}}(V_i-u_i)\cdot(V_k-u_k)\cdot f\cdot d^3\vec{V}\right)\cdot\frac{\partial u_k}{\partial t}+$$

$$+\frac{\partial}{\partial x_k}\left(\frac{\rho}{2}\cdot\int_{\vec{V}}(V_i-u_i)\cdot(V_k-u_k)\cdot(\vec{V}-\vec{u})^2\cdot f\cdot d^3\vec{V}\right)+\frac{\partial}{\partial x_k}\left(u_k\cdot\frac{\rho}{2}\cdot\int_{\vec{V}}(V_i-u_i)\cdot(\vec{V}-\vec{u})^2\cdot f\cdot d^3\vec{V}\right)+ \quad\text{(A4)}$$

$$+\left(\frac{\rho}{2}\cdot\int_{\vec{V}}(V_k-u_k)\cdot(\vec{V}-\vec{u})^2\cdot f\cdot d^3\vec{V}\right)\cdot\frac{\partial u_i}{\partial x_k}+\left(u_k\cdot\frac{\rho}{2}\cdot\int_{\vec{V}}(\vec{V}-\vec{u})^2\cdot f\cdot d^3\vec{V}\right)\cdot\frac{\partial u_i}{\partial x_k}+$$

$$+\left(\rho\cdot\int_{\vec{V}}(V_i-u_i)\cdot(V_k-u_k)\cdot(V_j-u_j)\cdot f\cdot d^3\vec{V}\right)\cdot\frac{\partial u_j}{\partial x_k}+\left(u_k\cdot\rho\cdot\int_{\vec{V}}(V_i-u_i)\cdot(V_j-u_j)\cdot f\cdot d^3\vec{V}\right)\cdot\frac{\partial u_j}{\partial x_k}=$$

$$=\frac{\rho}{2}\cdot\int_{\vec{V}}(V_i-u_i)\cdot(\vec{V}-\vec{u})^2\cdot St(f)\cdot d^3\vec{V}.$$

Expressing the third and ninth terms in the left-hand side of Eq. (A4) as

$$\left(\int_{\vec{V}}(V_i-u_i)\cdot(V_k-u_k)\cdot f\cdot d^3\vec{V}\right)\cdot\frac{\partial u_k}{\partial x_k}=\left(\int_{\vec{V}}\left((V_i-u_i)\cdot(V_k-u_k)-\delta_{ik}\cdot\frac{V_T^2}{2}\right)\cdot f\cdot d^3\vec{V}\right)\cdot\frac{\partial u_k}{\partial t}+\left(\frac{V_T^2}{2}\right)\cdot\frac{\partial u_k}{\partial t}$$

$$\left(u_k\cdot\int_{\vec{V}}(V_i-u_i)\cdot(V_j-u_j)\cdot f\cdot d^3\vec{V}\right)\cdot\frac{\partial u_j}{\partial x_k}=\left(u_k\cdot\int_{\vec{V}}\left((V_i-u_i)\cdot(V_j-u_j)-\delta_{ij}\cdot\frac{V_T^2}{2}\right)\cdot f\cdot d^3\vec{V}\right)\cdot\frac{\partial u_j}{\partial x_k}+\left(u_k\cdot\frac{V_T^2}{2}\right)\cdot\frac{\partial u_i}{\partial x_k}$$

and taking into account Eqs. (3) and the fact that $\sigma_{ij}=\sigma_{ji}$, we obtain

$$\frac{\partial q_i}{\partial t}+\left(\frac{5}{2}\cdot\frac{\rho\cdot V_T^2}{2}\right)\cdot\frac{\partial u_i}{\partial t}+\sigma_{ik}\cdot\frac{\partial u_k}{\partial t}+\frac{\partial}{\partial x_k}(u_k\cdot q_i)+q_k\cdot\frac{\partial u_i}{\partial x_k}+\left(u_k\cdot\frac{5}{2}\cdot\frac{\rho\cdot V_T^2}{2}\right)\cdot\frac{\partial u_i}{\partial x_k}+\left(u_k\cdot\sigma_{ij}\right)\cdot\frac{\partial u_j}{\partial x_k}+$$

$$+\frac{\partial}{\partial x_k}\left(\frac{\rho}{2}\cdot\int_{\vec{V}}(V_i-u_i)\cdot(\vec{V}-\vec{u})^2\cdot(V_k-u_k)\cdot f\cdot d^3\vec{V}\right)+\left(\rho\cdot\int_{\vec{V}}(V_i-u_i)\cdot(V_k-u_k)\cdot(V_j-u_j)\cdot f\cdot d^3\vec{V}\right)\cdot\frac{\partial u_j}{\partial x_k}\ =\ \text{(A5)}$$

$$=\frac{\rho}{2}\cdot\int_{\vec{V}}(V_i-u_i)\cdot(\vec{V}-\vec{u})^2\cdot St(f)\cdot d^3\vec{V},$$

where $\sigma_{zz}$ is given by Eq. (12). Substituting $\partial u_k/\partial t$ from Eq. (7) into Eq. (A5), and after some simple algebra, the equation for the heat flux $q_i$ turns into

$$\frac{\partial q_i}{\partial t} + \frac{\partial}{\partial x_k}(u_k \cdot q_i) + q_k \cdot \frac{\partial u_i}{\partial x_k} - \left(\frac{5}{2} \cdot \frac{V_T^2}{2}\right) \cdot \left(\frac{\partial}{\partial x_i}\left(\frac{\rho \cdot V_T^2}{2}\right) + \frac{\partial \sigma_{ji}}{\partial x_j}\right) - \frac{\sigma_{ik}}{\rho} \cdot \left(\frac{\partial}{\partial x_k}\left(\frac{\rho \cdot V_T^2}{2}\right) + \frac{\partial \sigma_{jk}}{\partial x_j}\right) +$$

$$+ \frac{\partial}{\partial x_k}\left(\frac{\rho}{2} \cdot \int_{\vec{V}} (V_i - u_i) \cdot (\vec{V} - \vec{u})^2 \cdot (V_k - u_k) \cdot f \cdot d^3\vec{V}\right) + \left(\rho \cdot \int_{\vec{V}} (V_i - u_i) \cdot (V_k - u_k) \cdot (V_j - u_j) \cdot f \cdot d^3\vec{V}\right) \cdot \frac{\partial u_j}{\partial x_k} = \quad (A6)$$

$$= \frac{\rho}{2} \cdot \int_{\vec{V}} (V_i - u_i) \cdot (\vec{V} - \vec{u})^2 \cdot St(f) \,.$$

APPENDIX B

Let us obtain equations for $M_{1x1y1z}$, Eq. (69). Multiplying the Boltzmann Equation, Eq. (1), with the BGK approximation of the collision term, Eq. (13), by

$$m \cdot 8 \cdot (V_x - u_x) \cdot (V_y - u_y) \cdot (V_z - u_z) \quad (B1)$$

we obtain

$$8 \cdot m \cdot (V_x - u_x) \cdot (V_y - u_y) \cdot (V_z - u_z) \cdot \frac{\partial (n \cdot f)}{\partial t} + 8 \cdot m \cdot (V_x - u_x) \cdot (V_y - u_y) \cdot (V_z - u_z) \cdot V_k \cdot \frac{\partial (n \cdot f)}{\partial x_k} =$$

$$= 8 \cdot m \cdot \tau^{-1} \cdot (V_x - u_x) \cdot (V_y - u_y) \cdot (V_z - u_z) \cdot (f - f_M) \,. \quad (B2)$$

Transferring the terms in front of the derivatives inside of the derivative brackets in Eq. (B2), we obtain

$$\frac{\partial}{\partial t}\left(8 \cdot \rho \cdot (V_x - u_x) \cdot (V_y - u_y) \cdot (V_z - u_z) \cdot f\right) + \left(8 \cdot \rho \cdot (V_y - u_y) \cdot (V_z - u_z) \cdot f\right) \cdot \frac{\partial u_x}{\partial t} +$$

$$+ \left(8 \cdot \rho \cdot (V_x - u_x) \cdot (V_z - u_z) \cdot f\right) \cdot \frac{\partial u_y}{\partial t} + \left(8 \cdot \rho \cdot (V_x - u_x) \cdot (V_y - u_y) \cdot f\right) \cdot \frac{\partial u_z}{\partial t} +$$

$$+ \frac{\partial}{\partial x_k}\left(8 \cdot \rho \cdot (V_k - u_k) \cdot (V_x - u_x) \cdot (V_y - u_y) \cdot (V_z - u_z) \cdot f\right) + \frac{\partial}{\partial x_k}\left(u_k \cdot 8 \cdot \rho \cdot (V_x - u_x) \cdot (V_y - u_y) \cdot (V_z - u_z) \cdot f\right) +$$

$$+ \left(8 \cdot \rho \cdot (V_k - u_k) \cdot (V_y - u_y) \cdot (V_z - u_z) \cdot f\right) \cdot \frac{\partial u_x}{\partial x_k} + \left(8 \cdot u_k \cdot \rho \cdot (V_y - u_y) \cdot (V_z - u_z) \cdot f\right) \cdot \frac{\partial u_x}{\partial x_k} +$$

$$+ \left(8 \cdot \rho \cdot (V_k - u_k) \cdot (V_x - u_x) \cdot (V_z - u_z) \cdot f\right) \cdot \frac{\partial u_y}{\partial x_k} + \left(8 \cdot u_k \cdot \rho \cdot (V_x - u_x) \cdot (V_z - u_z) \cdot f\right) \cdot \frac{\partial u_y}{\partial x_k} +$$

$$+ \left(8 \cdot \rho \cdot (V_k - u_k) \cdot (V_x - u_x) \cdot (V_y - u_y) \cdot f\right) \cdot \frac{\partial u_z}{\partial x_k} + \left(8 \cdot u_k \cdot \rho \cdot (V_x - u_x) \cdot (V_y - u_y) \cdot f\right) \cdot \frac{\partial u_z}{\partial x_k} =$$

$$= 8 \cdot m \cdot \tau^{-1} \cdot (V_x - u_x) \cdot (V_y - u_y) \cdot (V_z - u_z) \cdot (f - f_M) \,.$$

Substituting Eqs. (63) into Eq. (B3) and integrating the obtained equation over the entire velocity domain, we obtain

$$\frac{\partial}{\partial t}\left(8\cdot\rho\cdot V_T^3\cdot\int_{\vec{\chi}}\chi_x\cdot\chi_y\cdot\chi_z\cdot(f_{GRAD}+\tau\cdot f_M\cdot f_1)\cdot d^3\vec{\chi}^3\right)+\left(8\cdot\rho\cdot V_T^2\cdot\int_{\vec{\chi}}\chi_y\cdot\chi_z\cdot(f_{GRAD}+\tau\cdot f_M\cdot f_1)\cdot d^3\vec{\chi}^3\right)\cdot\frac{\partial u_x}{\partial t}+$$

$$+\left(8\cdot\rho\cdot V_T^2\cdot\int_{\vec{\chi}}\chi_x\cdot\chi_z\cdot(f_{GRAD}+\tau\cdot f_M\cdot f_1)\cdot d^3\vec{\chi}^3\right)\cdot\frac{\partial u_y}{\partial t}+\left(8\cdot\rho\cdot V_T^2\cdot\int_{\vec{\chi}}\chi_x\cdot\chi_y\cdot(f_{GRAD}+\tau\cdot f_M\cdot f_1)\cdot d^3\vec{\chi}^3\right)\cdot\frac{\partial u_z}{\partial t}+$$

$$+\frac{\partial}{\partial x_k}\left(8\cdot\rho\cdot V_T^4\cdot\int_{\vec{\chi}}\chi_k\cdot\chi_x\cdot\chi_y\cdot\chi_z\cdot(f_{GRAD}+\tau\cdot f_M\cdot f_1)\cdot d^3\vec{\chi}^3\right)+$$

$$+\frac{\partial}{\partial x_k}\left(u_k\cdot 8\cdot\rho\cdot V_T^3\cdot\int_{\vec{\chi}}\chi_x\cdot\chi_y\cdot\chi_z\cdot(f_{GRAD}+\tau\cdot f_M\cdot f_1)\cdot d^3\vec{\chi}^3\right)+$$

$$+\left(8\cdot\rho\cdot V_T^3\cdot\int_{\vec{\chi}}\chi_k\cdot\chi_y\cdot\chi_z\cdot(f_{GRAD}+\tau\cdot f_M\cdot f_1)\cdot d^3\vec{\chi}^3\right)\cdot\frac{\partial u_x}{\partial x_k}+\left(u_k\cdot 8\cdot\rho\cdot V_T^2\cdot\int_{\vec{\chi}}\chi_y\cdot\chi_z\cdot(f_{GRAD}+\tau\cdot f_M\cdot f_1)\cdot d^3\vec{\chi}^3\right)\cdot\frac{\partial u_x}{\partial x_k}+$$

$$+\left(8\cdot\rho\cdot V_T^3\cdot\int_{\vec{\chi}}\chi_k\cdot\chi_x\cdot\chi_z\cdot(f_{GRAD}+\tau\cdot f_M\cdot f_1)\cdot d^3\vec{\chi}^3\right)\cdot\frac{\partial u_y}{\partial x_k}+\left(u_k\cdot 8\cdot\rho\cdot V_T^2\cdot\int_{\vec{\chi}}\chi_x\cdot\chi_z\cdot(f_{GRAD}+\tau\cdot f_M\cdot f_1)\cdot d^3\vec{\chi}^3\right)\cdot\frac{\partial u_y}{\partial x_k}+$$

$$+\left(8\cdot\rho\cdot V_T^3\cdot\int_{\vec{\chi}}\chi_k\cdot\chi_x\cdot\chi_y\cdot(f_{GRAD}+\tau\cdot f_M\cdot f_1)\cdot d^3\vec{\chi}^3\right)\cdot\frac{\partial u_z}{\partial x_k}+\left(u_k\cdot 8\cdot\rho\cdot V_T^2\cdot\int_{\vec{\chi}}\chi_x\cdot\chi_y\cdot(f_{GRAD}+\tau\cdot f_M\cdot f_1)\cdot d^3\vec{\chi}^3\right)\cdot\frac{\partial u_z}{\partial x_k}=$$

$$=-8\cdot\rho\cdot V_T^3\cdot\int_{\vec{\chi}}\chi_x\cdot\chi_y\cdot\chi_z\cdot f_M\cdot f_1\cdot d^3\vec{\chi}^3.$$

(B4)

Following the Chapman-Enskog recipe [5, 6], let us put $\tau\to 0$ into Eq. (B4). Then substituting Eq. (36) for $f_{GRAD}$ in the left-hand side of the obtained equation we obtain the following equation for $M_{1x1y1z}$:

$$-\frac{M_{1x1y1z}}{\tau}=-8\cdot\rho\cdot V_T^3\cdot\int_{\vec{\chi}}\chi_x\cdot\chi_y\cdot\chi_z\cdot f_M\cdot f_1\cdot d^3\vec{\chi}=(8\cdot\sigma_{yz})\cdot\frac{\partial u_x}{\partial t}+(8\cdot\sigma_{xz})\cdot\frac{\partial u_y}{\partial t}+(8\cdot\sigma_{xy})\cdot\frac{\partial u_z}{\partial t}+$$

$$+(8\cdot\sigma_{yz}\cdot u_k)\cdot\frac{\partial u_x}{\partial x_k}+(\sigma_{xz}\cdot u_k)\cdot\frac{\partial u_y}{\partial x_k}+(8\cdot\sigma_{xy}\cdot u_k)\cdot\frac{\partial u_z}{\partial x_k}+\left(\frac{16\cdot q_z}{5}\right)\cdot\left(\frac{\partial u_x}{\partial y}+\frac{\partial u_y}{\partial x}\right)+\left(\frac{16\cdot q_y}{5}\right)\cdot\left(\frac{\partial u_x}{\partial z}+\frac{\partial u_z}{\partial x}\right)+$$

$$+\left(\frac{16\cdot q_x}{5}\right)\cdot\left(\frac{\partial u_y}{\partial z}+\frac{\partial u_z}{\partial y}\right)+\frac{\partial}{\partial x}\left(4\cdot V_T^2\cdot\sigma_{yz}\right)+\frac{\partial}{\partial y}\left(4\cdot V_T^2\cdot\sigma_{xz}\right)+\frac{\partial}{\partial z}\left(4\cdot V_T^2\cdot\sigma_{xy}\right).$$

(B5)

By substituting the expression for $\partial u_i / \partial t$, Eq. (7), into Eq. (B5) we obtain the final expression for $M_{1x1y1z}$

$$-\frac{M_{1x1y1z}}{\tau} = 4 \cdot \rho \cdot V_T^2 \cdot \left[ \frac{\partial}{\partial x}\left(\frac{\sigma_{yz}}{\rho}\right) + \frac{\partial}{\partial y}\left(\frac{\sigma_{xz}}{\rho}\right) + \frac{\partial}{\partial z}\left(\frac{\sigma_{xy}}{\rho}\right) \right] - \left(\frac{8 \cdot \sigma_{yz}}{\rho}\right) \cdot \frac{\partial \sigma_{kx}}{\partial x_k} - \left(\frac{8 \cdot \sigma_{xz}}{\rho}\right) \cdot \frac{\partial \sigma_{ky}}{\partial x_k} - \left(\frac{8 \cdot \sigma_{xy}}{\rho}\right) \cdot \frac{\partial \sigma_{kz}}{\partial x_k} + \left(\frac{16 \cdot q_z}{5}\right) \cdot \left(\frac{\partial u_x}{\partial y} + \frac{\partial u_y}{\partial x}\right) + \left(\frac{16 \cdot q_y}{5}\right) \cdot \left(\frac{\partial u_x}{\partial z} + \frac{\partial u_z}{\partial x}\right) + \left(\frac{16 \cdot q_x}{5}\right) \cdot \left(\frac{\partial u_y}{\partial z} + \frac{\partial u_z}{\partial y}\right). \tag{B6}$$

Taking into account

$$\sigma_{ij} \propto \tau \cdot \frac{\rho \cdot V_T^2 \cdot u}{L} \quad \text{and} \quad q_i \propto \tau \cdot \frac{\rho \cdot V_T^4}{L}, \tag{B7}$$

we can drop terms that are of the order $\tau^2$ in the right-hand side of Eq. (B6); this yields

$$-M_{1x1y1z} = -\tau \cdot \left\{ 4 \cdot \rho \cdot V_T^2 \cdot \left[ \frac{\partial}{\partial x}\left(\frac{\sigma_{yz}}{\rho}\right) + \frac{\partial}{\partial y}\left(\frac{\sigma_{xz}}{\rho}\right) + \frac{\partial}{\partial z}\left(\frac{\sigma_{xy}}{\rho}\right) \right] + \left(\frac{16 \cdot q_z}{5}\right) \cdot \left(\frac{\partial u_x}{\partial y} + \frac{\partial u_y}{\partial x}\right) + \left(\frac{16 \cdot q_y}{5}\right) \cdot \left(\frac{\partial u_x}{\partial z} + \frac{\partial u_z}{\partial x}\right) + \left(\frac{16 \cdot q_x}{5}\right) \cdot \left(\frac{\partial u_y}{\partial z} + \frac{\partial u_z}{\partial y}\right) \right\}. \tag{B8}$$

Multiplying the Boltzmann Equation, Eq. (1), with the BGK approximation of the collision term, Eq. (13), by

$$m \cdot \left(4 \cdot (V_x - u_x)^2 - 2 \cdot V_T^2\right) \cdot \left(4 \cdot (V_y - u_y)^2 - 2 \cdot V_T^2\right), \tag{B9}$$

$$m \cdot \left(16 \cdot (V_x - u_x)^4 - 48 \cdot V_T^2 \cdot (V_x - u_x)^2 + 12 \cdot V_T^4\right), \tag{B10}$$

$$8 \cdot m \cdot \left(2 \cdot (V_x - u_x)^3 - 3 \cdot V_T^2 \cdot (V_x - u_x)\right) \cdot (V_y - u_y), \tag{B11}$$

$$8 \cdot m \cdot (V_x - u_x) \cdot (V_y - u_y) \cdot \left(2 \cdot (V_z - u_z)^2 - V_T^2\right), \tag{B12}$$

$$m \cdot \left(8 \cdot (V_x - u_x)^3 - 12 \cdot (V_x - u_x) \cdot \left((V_y - u_y)^2 + (V_y - u_y)^2\right)\right), \tag{B13}$$

$$8 \cdot m \cdot (V_x - u_x) \cdot \left((V_y - u_y)^2 - (V_z - u_z)^2\right), \tag{B14}$$

and following the same procedure as for the Moment $M_{1x1y1z}$, we obtain equations for $M_{2x2y}$, $M_{4x}$, $M_{3x1y}$, $M_{1x1y2z}$, $M_{x(2y+2z)}$, $M_{1x(2y-2z)}$ respectively.

ACKNOWLEDGEMENTS

The authors would like to express their gratitude to ERC Incorporation for support of this project, D. Pekker and A. Alekseenko for helpful discussions, and S. Keitz, A. Pekker, and Y. Pekker for their kind help in preparing the text of this paper.